\documentclass[numberedappendix]{emulateapj}
\usepackage{graphicx}
\usepackage[space]{grffile}
\usepackage{latexsym}
\usepackage{textcomp}
\usepackage{longtable}
\usepackage{multirow,booktabs}
\usepackage{amsfonts,amsmath,amssymb}
\usepackage{natbib}
\usepackage{url}
\usepackage{hyperref}
\hypersetup{colorlinks=false,pdfborder={0 0 0}}
\newif\iflatexml\latexmlfalse
\usepackage[utf8]{inputenc}
\usepackage[english]{babel}

\newcommand{\msun}{M_\odot}
\newcommand{\rxte}{{\it RXTE}}

\newcommand{\gx}{{GX~339--4}}

\newcommand{\mdot}{\dot{M}}

\newcommand{\xmm}{{\it XMM-Newton}}

\newcommand{\lph}{L_{\rm ph}}

\newcommand{\spin}{a_*}

\newcommand{\simpl}{{\sc simpl}}
\newcommand{\simplcut}{{\sc simplcut}}

\newcommand{\rin}{R_{\rm in}}
\newcommand{\risco}{R_{\rm ISCO}}
\newcommand{\Risco}{R_{\rm ISCO}}

\newcommand{\chisq}{\chi^{2}}

\newcommand{\fsc}{f_{\rm SC}}

\newcommand{\keV}{\rm keV}

\newcommand{\cm}{\rm cm}

\newcommand{\ecut}{E_{\rm cut}}
\newcommand{\Ecut}{E_{\rm cut}}
\newcommand{\kTe}{kT_e}
\newcommand{\kte}{kT_e}
\newcommand{\nthcomp}{{\sc nthcomp}}
  
\newcommand{\relxill}{{\sc relxill}}
\newcommand{\xillver}{{\sc xillver}}

\newcommand{\xspec}{{\sc xspec}}

\newcommand{\RF}{R_{\rm F}}
\newcommand{\rf}{R_{\rm F}}

\newcommand{\nustar}{{\em NuSTAR}}

\usepackage{color}

\begin{document}
\title{Self-Consistent Black Hole Accretion Spectral Models and the Forgotten Role of Coronal Comptonization of Reflection Emission}

\shorttitle{Compton Scattering of BH Disk Reflection}
                
\shortauthors{Steiner et al.}

\author{James F.\ Steiner\altaffilmark{1}\altaffilmark{\dag}, Javier A.\ Garc{\'{\i}}a\altaffilmark{2,3,4}\altaffilmark{\ddag}, \\ 
	Wiebke Eikmann\altaffilmark{4}, Jeffrey E.\ McClintock\altaffilmark{3}, Laura W.\ Brenneman\altaffilmark{3}, Thomas Dauser\altaffilmark{4},  Andrew C.\ Fabian\altaffilmark{5}}

	\altaffiltext{\dag}{Einstein Fellow.}
	\altaffiltext{\ddag}{Alexander von Humboldt Fellow.}
\altaffiltext{1}{MIT Kavli Institute for Astrophysics and Space  Research, MIT, 70 Vassar Street, Cambridge, MA 02139.}
\altaffiltext{2}{Cahill Center for Astronomy and Astrophysics, California Institute of Technology, Pasadena, CA 91125.}
\altaffiltext{3}{Harvard-Smithsonian Center for Astrophysics, 60  Garden Street, Cambridge, MA 02138.}
\altaffiltext{4}{Remeis Observatory \& ECAP, Universit\"at Erlangen-N\"urnberg, Sternwartstr.~7, 96049 Bamberg, Germany.} 
\altaffiltext{5}{Department of Astronomy, Cambridge University,
  Madingley Road, Cambridge, CB3 0HA, UK.}
  
\email{jsteiner@mit.edu}
\begin{abstract}

Continuum and reflection spectral models have each been widely  employed in measuring the spins of accreting black holes.  However, the two approaches have not been implemented together in a  photon-conserving, self-consistent framework.  We develop such a framework using the black-hole X-ray binary GX~ 339--4 as a touchstone source, and we demonstrate three important ramifications: (1) Compton scattering of reflection emission in the corona  is routinely ignored, but is an essential consideration  given that reflection is linked to the regimes with strongest Comptonization.  Properly accounting for this causes the inferred reflection fraction to increase substantially, especially for the hard state.   Another  important impact of the Comptonization of reflection emission by the corona is the downscattered tail.   Downscattering has the potential to mimic the  relativistically-broadened red wing of the Fe line associated with a spinning black hole. (2) Recent evidence for a reflection component with a   harder spectral index than the power-law continuum is naturally explained as Compton-scattered reflection emission. (3) Photon conservation provides an important constraint on the hard state's accretion rate.   For {\em bright} hard states, we show that disk truncation to large scales $R >> R_{\rm ISCO}$ is unlikely as this would require accretion rates far in excess of the observed $\mdot$ of the brightest soft states.  Our principal conclusion is that when modeling relativistically-broadened reflection, spectral models should allow for coronal Compton scattering of the reflection features, and when possible, take advantage of the additional constraining power from linking to the thermal disk component.

\end{abstract}\label{abstract}
  

\section{Introduction}\label{section:intro}

Nature's black holes are delineated by mass into two primary classes:
supermassive ($M \gtrsim 10^5 \msun$) and stellar mass ($M \lesssim 100
\msun$), with any falling in-between termed {\em intermediate} mass.
For a black hole of any mass, the no-hair theorem holds that the black
hole is uniquely and completely characterized by its mass and its spin
angular momentum.  One of the principal challenges in modern
astrophysics is to measure and understand the distribution of black hole
spins ($\spin$).

One of the most widely-employed approaches for measuring black hole spin
is through modeling fluoresced features in the {\em reflection}
spectrum.  The most prominent and recognized such feature is the
Fe-K$\alpha$ line complex \citep{Fabian_1989, Brenneman_Reynolds,
  Brenneman_2013, Reynolds_2014}.  The term ``reflection'' here refers
to the reprocessing of hard X-ray emission illuminating the disk from
above.  The illuminating agent is understood to be a hot corona
enshrouding the inner disk, and the hard X-ray emission is attributed to
Compton up-scattering of the thermal seed photons originating in the
disk component (e.g., \citealt{White_1982, Gierlinski_1999}).  The
Compton emission from the corona has a power-law spectrum, which
(generally) cuts off at high energies.  A portion of this coronal
emission returns to the disk and irradiates its outer atmosphere. The
disk's heated outer layer is photoionized, resulting in fluorescent line
emission as the atoms de-excite. In addition to its forest of line
features, the reflection continuum produces a broad ``Compton hump'' peaked at
$\sim 30$~keV, above the Fe edge.

Relativity sets Doppler splitting and boosting as well as
gravitational redshift for the reflection features produced across the
disk. Each such feature is accordingly imprinted with information about
the spacetime at its point of origin in the disk. A feature of principal
interest is the Fe K line whose red wing is used to estimate the disk's
inner radius, which in turn provides a constraint on the black hole's
spin.

The other primary method for measuring black hole spin is via thermal
continuum-fitting \citep{Zhang_1997, MNS14}. In this case, the
blackbody-like emission from the accretion disk is the component of
interest, and spin manifests through the efficiency with which the disk radiates
away the rest-mass energy of accreting gas.
In effect, the spin is estimated using the combined constraint
provided by the disk's observed flux and peak temperature.  For this
method to deliver an estimate of spin, it is necessary to have knowledge
of the black hole's mass, the line-of-sight inclination of the spin
axis, and the system's distance.

The reflection method is applied to both stellar-mass and supermassive
black holes, whereas continuum fitting is predominantly useful for
stellar-mass black holes.  Both methods rely upon a single crucial
foundational assumption: that the inner edge of the accretion disk is
exactly matched to the radius of the innermost-stable circular orbit
(ISCO), which is a monotonic function of both mass and spin.

Although there is overlap in the set of BH systems which have measurements from each approach,  the methods are optimized for opposing conditions; when a system is most amenable for one method, the other is
generally hampered.  For instance, for the thermal state in which
continuum-fitting is most adept (e.g., \citealt{MNS14,
  Steiner_2009}), Comptonization and reflection
are faint compared to the bright disk. Conversely,
hard states are dominated by the Compton power law and its associated
reflection (e.g., \citealt{Fabian_Ross}), and here the thermal disk is quite
cool and faint, often so weak that the thermal emission is undetected.  As a result, both methods are not usually applied fruitfully to a single data set
(see, e.g., \citealt{Steiner_lmcx1}). Instead, for transient black-hole
X-ray binaries, one can apply both methods to observations at distinct epochs
capturing a range of hard and soft states as the source evolves.

As consequence of this phenomenological decoupling between thermal and nonthermal dominance, the spectral models describing thermal disk emission versus Compton and reflection emission have largely undergone
independent development.  
And as a result the cross-coupling between spectral components has been sparsely studied.  One notable effort by \citet{Petrucci_2001} explored the effect of coronal Comptonization on the reflection continuum flux and Compton hump, but in general this has gone unexplored.  
A handful of similar efforts have taken steps towards self-consistent treatment of the thermal and nonthermal spectral components 
-- notably using {\sc eqpair} (\citealt{EQPAIR}; e.g.,
\citealt{Kubota_2016}) or {\sc compps} \citep{compps}, and to lesser
degree \citet{Parker_gx2016,
  Basak_Zdziarski, Plant_2015, Tomsick_2014, Steiner_j1550spin, Miller_2009}. But an advanced self-consistent approach has not been realized. In this work, we examine several
important aspects of a self-consistent model, sighted towards measuring
black hole spin.

Adopting the usual assumption that the hard power-law photons originate
in a hot corona that Compton scatters thermal disk photons, then by
extension the reflection photons emerging from the inner disk will likewise undergo the same coronal Compton scattering.  Recognizing this, as a first step towards
self-consistency, we begin by examining the impact of coronal Comptonization on
the reflection spectrum.  This is particularly important for the hard
state, in which Compton scattering is most pronounced.  After touching upon this point, we go on to tackle the larger objective of
producing an interlinked disk-coronal spectral model which imposes self-consistency.  To connect with
observations, we apply our model to the peak bright hard-state spectrum
of \gx, which is described in detail in \citet{Garcia_GX}.

These data are among the highest-signal reflection spectra
ever studied. Specifically, the data in question correspond to ``Box A'' in
\citet[][hereafter,G15]{Garcia_GX}, which is comprised of 
$>40$-million X-ray counts as collected by \rxte's Proportional Counter Array (PCA; \citealt{PCA}) with a spectral range $\sim3-45$~keV.

In G15, a simple spectral model was adopted, consisting of a cutoff power-law continuum, a component of relativistically broadened
reflection ({\sc relxill}), and a narrow component of distant
reflection ({\sc xillver}; i.e., located far from the regime of strong
gravity and also far from the corona).  In addition, a cosmetic Gaussian absorption feature was
included in the model near the Fe complex at $\sim 7.4$~keV.  For further details on the data and modeling
procedure adopted, we refer the reader to G15.

G15 followed the common practices standard in reflection modeling of
hard-state data, which do not include the self-consistent effects we
introduce here.  In G15, the reflection fits were found to clearly
demand that reflection originate from a very small inner radius, a
radius of $\lesssim 2~R_g$.  And yet no thermal emission was
required by the fit, although two reflection components were demanded (the broad and narrow components mentioned above).  Given the high luminosity of this hard state, can the reflection fit  be reconciled
with the apparent dearth of thermal disk emission?  What is the effect
of including Compton-scattering on the reflection emission?  Can a
single reflection component, partially transmitted and partially
scattered by the corona, account for the complete signal?  Our work addresses these questions using the \gx\ data as a touchstone data for examining the impact of a self-consistent framework on black-hole spectral data as a general practice.

Section~\ref{section:model} describes our overall approach.  We first focus on the
impact of Compton-scattering on Fe lines in reflection spectra in
Section~\ref{section:reflectionsc}, and then discuss a fully self-consistent approach in Section~\ref{section:fullmodel}.  In
Section~\ref{section:results}, we apply this prescription to \gx.
Finally, a broader discussion and our conclusions are given in
Sections~\ref{section:discussion} and \ref{section:concs}.
	
\begin{figure*}[]
\begin{center}
  \includegraphics[width=2.125\columnwidth]{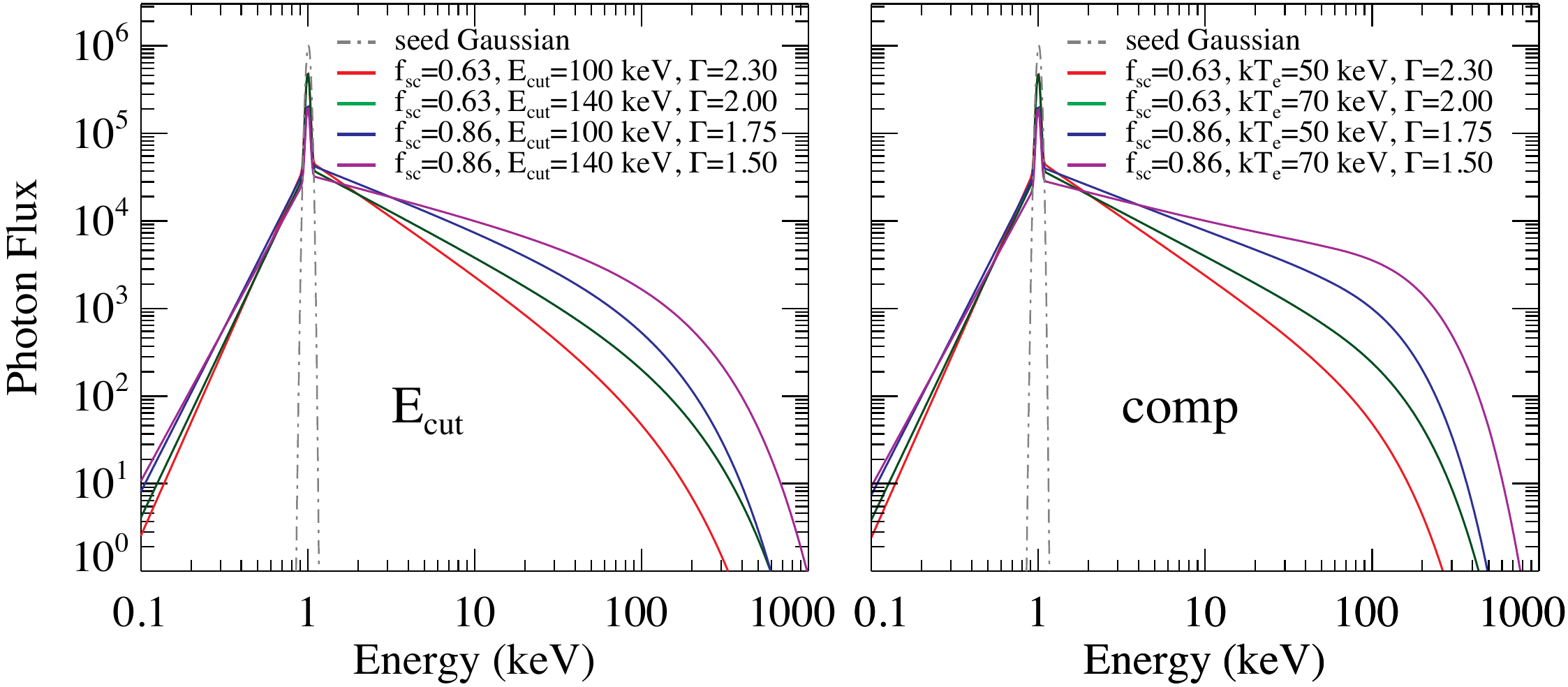}
  \vspace{-0.15cm}
  \caption{The shape of the Compton-scattered continuum that
    results from seed photons input at 1~keV, for a range of coronal
    properties.  We show the output of \simplcut\ acting on an input Gaussian line at 1~keV using each of its two kernels: $\ecut$ (left) which we adopt throughout the text, and {\em comp} (right) which is based on \nthcomp\ and described in Appendix~\ref{append:simplcomp}.
    The values of $\fsc$ correspond to $\tau=1$ and $\tau=2$ for uniform coronae with unity covering fraction. 
        Note that here {\em most} line photons go into the scattered wings; these appear faint with respect to the peak because they are very broad even though they contain most of the signal.
    Of principal importance is the
    significant downscattered contribution from the line, which is
    appreciable and is steeper when $\Gamma$ is large.  
    }\label{fig:comptonization}
\end{center}
\end{figure*}

\begin{figure}[]
	\begin{center}
	  \includegraphics[width=1\columnwidth]{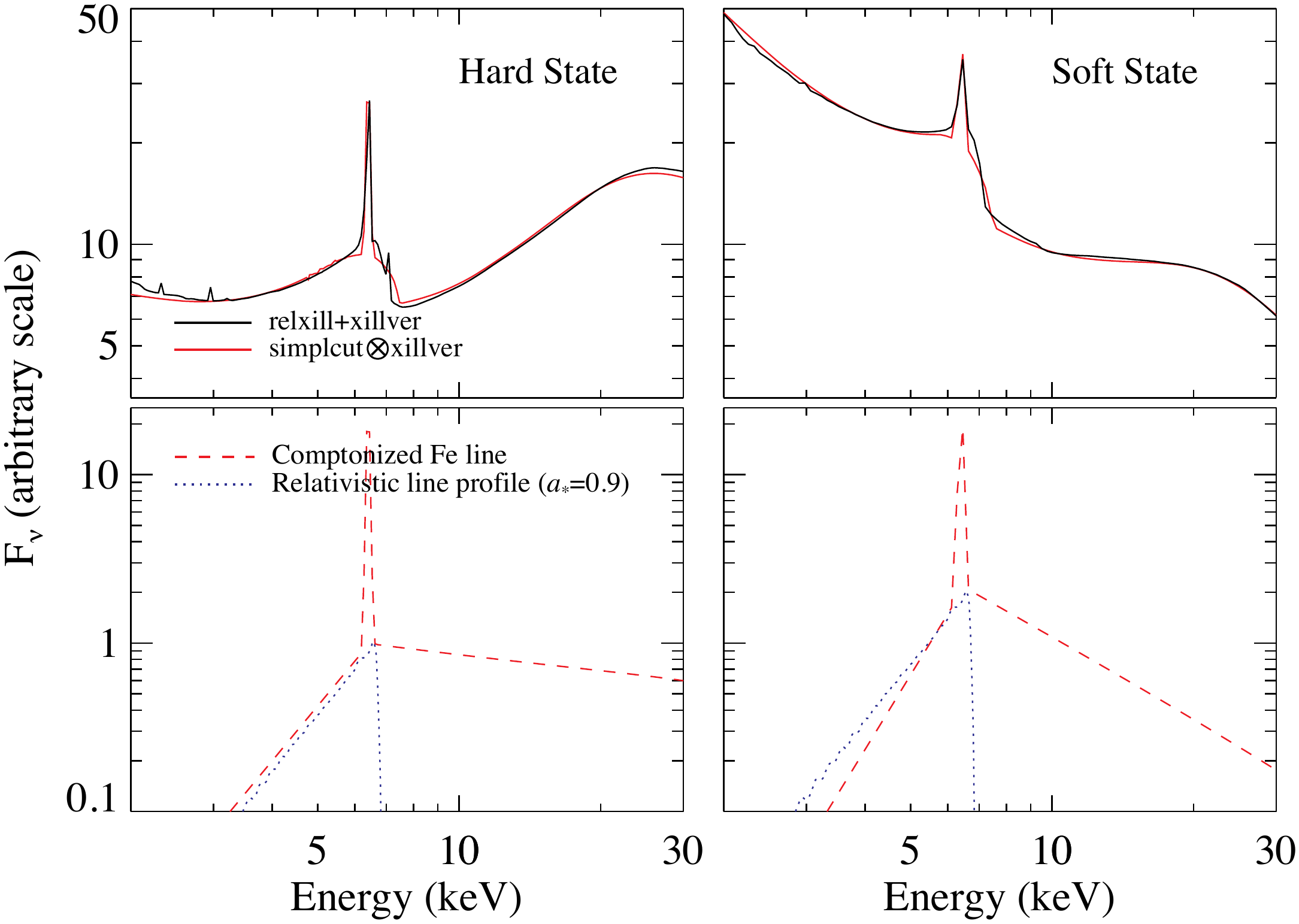}
          \caption{An illustration showing how downscattered reflection
            could plausibly mimic broad plus narrow reflection components and
            could therefore potentially be a source of systematic error
            in estimating spin.  Top panels depict composite spectra
            modeled as {\em either} broad and narrow reflection or as a
            Comptonized narrow reflection component.  The bottom
            panels present the profile of the modeled coronal scattering as acting on a narrow Fe line to illustrate its shape.    For
            reference, we also overlay the shape of a relativistically
            broadened line with spin $\spin=0.9$.  Note from the bottom panels that the broad relativistic line 
            profile matches closely that of the Compton-downscattered wing for the hard state (with a steeper
            $\Gamma=1.5$), but for the soft state ($\Gamma=2.5$) the red
            wing is narrower.}\label{fig:mimic}
	\end{center}
\end{figure}

\begin{figure}[]
	\begin{center}
	  \includegraphics[width=1\columnwidth]{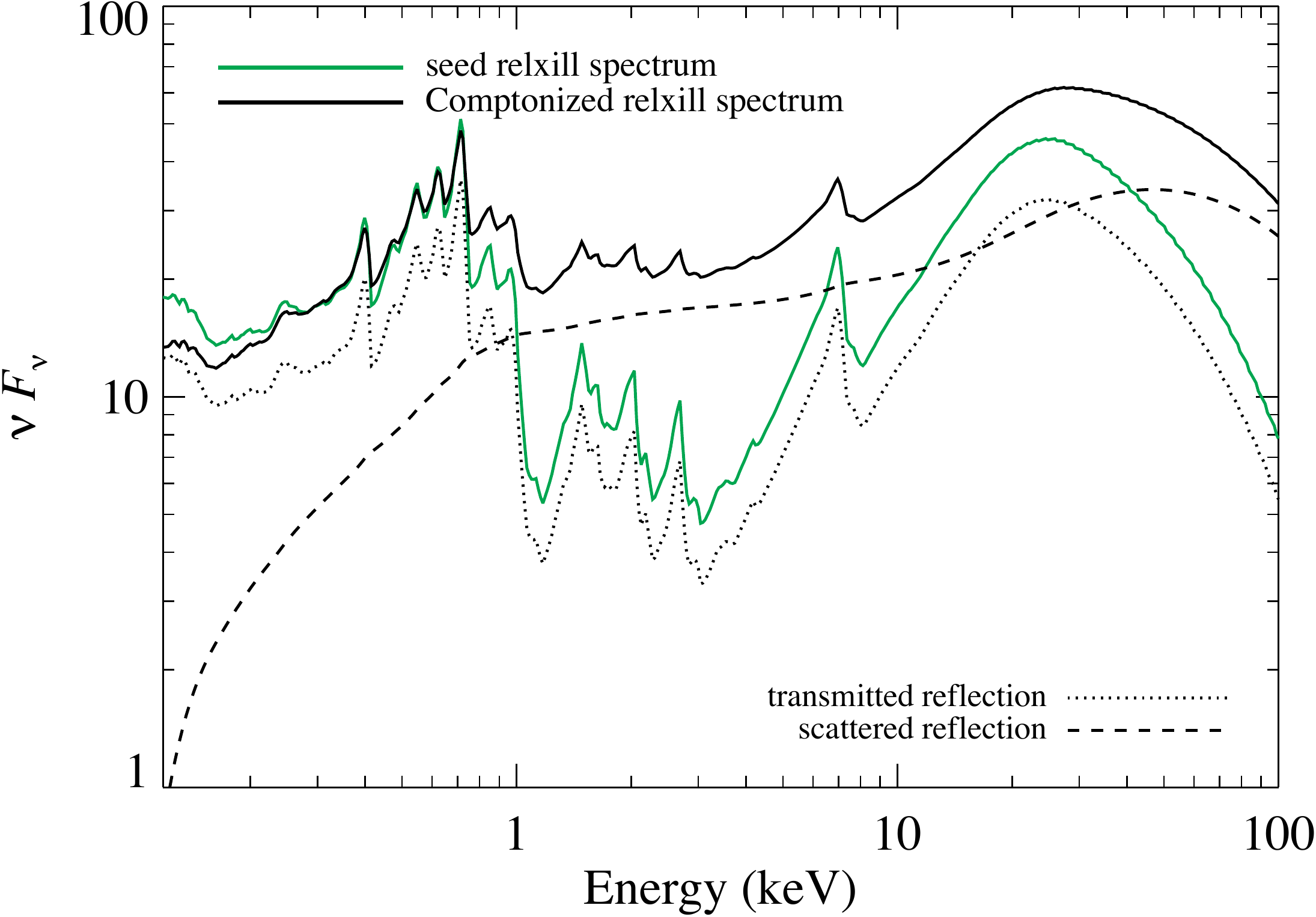}
          \caption{The effect of Compton scattering on a broad reflection component (\relxill) component.  The intrinsic emission (i.e., the reflection component for a completely transparent corona) is shown as a green solid line.  The solid black line shows the net effect of Compton-scattering on this emission.  Dashed and dotted black lines show the portions which have respectively scattered or been transmitted through the corona. 
          The spectrum illustrated here was generated for $\fsc=0.3$,  $\Gamma=2$, and $\ecut=100$~keV for the corona; $\spin=0$, $q=3$, $i=45\degr$, log~$\xi$=2, and a solar abundance of Fe for the disk.  The bolometric photon count is identical for both curves; the apparent enhancement for the black curve in the 1--100 keV range is due to the $\nu F_\nu$ scaling, and is compensated at the lowest energies ($\lesssim 1~\keV$) at which the green curve is brighter (and which dominates the photon count).}\label{fig:rxscat}
	\end{center}
\end{figure}

\section{Structuring a Self-Consistent Model}\label{section:model}

We focus on a self-consistent disk-coronal spectral model, in which the
thermal disk emission, reflection, and coronal power-law
are interconnected in a single framework.  Our approach makes several
simplifying assumptions.  We firstly assume that the power-law component
is generated by Comptonization in a thermal corona with no bulk motion, and that the
corona is uniform in the sense that its temperature, optical depth, and
covering factor\footnote{While it is often assumed that optical depth
  fully determines the fraction of emitted photons that scatter in the
  corona, this is only true for a fixed covering factor.} are
invariant across the inner disk (i.e., we ignore any radial gradients
that may affect the corona-disk interplay).   While we do not impose a particular geometry on the corona, we
  make the simplifying assumption that the corona emits isotropically. For a chosen geometry, an anisotropy correction (see e.g.,
\citealt{Haardt_1993}) could be applied to the reflection fraction; an investigation of such geometry-specific corrections is left for future work.

We assume that the
disk is optically thick and geometrically thin and that it terminates at
an inner radius $\rin \geq \risco$.  The boundary condition at the inner
radius assumes zero torque\footnote{\citet{Basak_Zdziarski} point out that the
  zero-torque condition may not be applicable when the disk is truncated
  at $\rin > \risco$.  A detailed consideration of the precise boundary
  torque is beyond our scope.}.  Spectral hardening (also commonly termed
the ``color correction'') describes the relationship between color
temperature $T_c$ and effective temperature $T_{\rm eff}$, and it is
defined as $T_c = f T_{\rm eff}$.  The correction term $f$ is assumed to
scale as $f \sim T_c^{1/3}$ \citep{bhspec}.  Radiation emitted by the
disk that is bent back and strikes the disk due to gravitational lensing (i.e.,
``returning radiation'') is not considered here.

A {\em fully} self-consistent model of the accreting system should include the following couplings between components:

\begin{itemize}
\item {\bf disk-corona}: A fraction $f_{\rm sc}$ of thermal accretion
  disk photons are Compton-scattered by the corona into the power-law
  component.  Reflection photons, which originate at the disk surface,
  will also scatter in the hot corona.  Photon conservation here ensures
  that the Compton power-law photons once
  originated as thermal disk emission.

\item {\bf corona-reflection}: The Compton-scattered photons which
  illuminate to the disk give rise to the reflection component.  The {\em
    reflection fraction} describes the flux of Compton-scattered photons
  directed back to the disk relative to the Compton flux that is
  transmitted to infinity \citep{Dauser_2016}.  Properly incorporating
  {\em photon counting} ensures that the Compton-scattered disk photons
  are conserved and correctly apportioned between components.  For
  instance, in this case photon conservation links the flux of the observed
  Compton power-law component to the flux of this same component from the vantage point of the disk (i.e., as prescribed by the reflection fraction).

\item {\bf disk-reflection}: The inner radius of the disk is tracked by
  two independent spectral characteristics: the red wing of the
  relativistic reflection features and the emitting area of the
  multicolor-blackbody disk. A self-consistent treatment of these two
  components is achieved by ensuring that the radius of the thermal
  emission in the soft state and the reflection component in the hard
  state mutually constrain the area of the thermal emission for the
  hard-state disk.  At the same time, photon conservation provides a
  constraint on the temperature via the disk photon luminosity.

\item {\bf corona/disk-jet}:  Although self-consistent modeling of the radio jet can constrain the hard-state geometry (see e.g., \citealt{Coriat_2011}), the radio jet is beyond our scope is not considered in this work.

\end{itemize}

  \section{Compton-Scattering of Reflection Photons}\label{section:reflectionsc}

  We begin our investigation of self-consistent modeling with an
  in-depth examination of the coupling of the
  thermal disk photons and the coronal electrons which give rise to the Compton
  power law. The Compton scattering of thermal disk photons into a
  power-law component has been widely studied (e.g., \citealt{compps, comptt,
    kyrmodels, nthcomp}, and references therein). However, the impact of
  Compton scattering on the {\em non-thermal} reflection features
  emitted from the disk's surface is typically ignored, though see
  \citet{Wilkins_2015} and \citet{Petrucci_2001}.

  As a first-order treatment, we can convolve the reflection spectrum
  with the Compton-scattering kernel such as \simpl\ \citep{Steiner_simpl}.
  This model redistributes a scattered fraction $\fsc$ of seed photons
  into a post-scattering distribution via a Green's function.  The most
  basic implementation consists of a one- or two-sided power law
  distribution \citep{Steiner_2009}. In this paper we adopt the
  two-sided version and go one step further, introducing a more sophisticated implementation of \simpl\ 
  termed \simplcut, which we will use throughout and which we now describe.

\subsection{simplcut}\label{subsec:simplcut} 

\simplcut\footnote{Available at
  \url{http://space.mit.edu/~jsteiner/simplcut.html}} is an extension of
the \simpl\ model that adopts a cut-off power-law shape for the Compton component.  It is governed by four physical parameters: the
scattered fraction $\fsc$, the spectral index $\Gamma$, the high-energy
turnover, and the reflection fraction ($\RF$).  We note that $\fsc$ is distinct from the optical depth $\tau$; $\fsc$, and $\Gamma$ are fitted, independent of $\tau$. This operationally allows for a variable (non-unity) covering factor of the corona above the disk.  ($\tau$ can be  inferred for a desired geometry using the spectral
 index and high-energy turnover terms.)
$\RF$ is defined as the
number of {\em scattered} photons which return to illuminate the disk
(i.e., those producing reflection) divided by the number which reach
infinity. No prescription for angle-dependence is applied, which is
equivalent to using the observed power-law flux as proxy for all flux at
infinity.  

Within \simplcut\ are two options for the scattering kernel.  The first kernel, which we adopt throughout unless otherwise specified, is shaped by an exponential cutoff $\Ecut$ and
described by the following Green's function, normalized at each seed
energy $E_0$:

\begin{align}\label{eqn:simplcut}
G(E;E_0)dE \propto \begin{cases}
(E/E_0)^{-\Gamma}~{\rm exp}(-E/\ecut) dE/E_0,  &E\geq E_{0} \\
(E/E_0)^{\Gamma+1} dE/E_0, &E < E_0,
\end{cases}
\end{align}

The second kernel is based upon \nthcomp\ \citep{nthcomp}, with electron temperature $\kTe$ taking the place of $\Ecut$.  This kernel and its implementation are described in Appendix~\ref{append:simplcomp}, and its shape is illustrated in the right panel of Figure~\ref{fig:comptonization}.  We opt to use the $\ecut$ kernel throughout for the sake of its straightforward comparison with published results (though see Appendix~\ref{append:simplcomp} to see our modeling results using both kernels).

Figure~\ref{fig:comptonization} demonstrates \simplcut's effect, illustrating the net impact of the corona on a narrow 1~keV line viewed through coronae with a range of settings.  
The gray lines depict seed 1~keV photons. For a covering factor of unity,
the familiar optical depth $\tau$ is related to the scattered fraction by $\fsc =
1-{\rm exp}(-\tau)$; electron temperature $\kte$ and high-energy cutoff $\ecut$ are approximately matched to one another using an approximate correspondence $\ecut=2-3~\kte$ (here we have arbitrarily selected $\ecut=2~\kte$ for presenting the kernels).

The most important feature to note is the prominent downscattering of
the line by the hot corona (e.g., \citealt{Matt_1997}), which is essentially identical between the two kernels.  This is worth particular attention given
that the red wing of reflection features are widely employed in
estimating the inner-disk radius (and thereby the spin) of accreting
black holes.  Note also that the transmitted
portion of the line remains narrow.
In considering the Compton-scattering of reflection features, our prescription using \simplcut\ greatly improves upon the present {\em status quo}, which is seen in the figure as the narrow gray line.

\subsection{On Compton Downscattered Line Features}\label{subsec:downscatter}

Given that Compton scattering produces emission that can contribute to
the red wing of a line profile (notably the Fe line), this may plausibly impact 
spin measurements.  We now explore this phenomenon in greater detail. The
principal question before us is whether or not Compton downscattering
can mimic the effect of strong gravity on fluorescent line emission.
This is an important consideration given that the hard states in which
reflection is widely studied is associated with strong Compton
scattering.

As described by Eqn.~\ref{eqn:simplcut}, the shape of the {\em
  downscattered} emission is purely a function of $\Gamma$
(specifically, the downscattering spectral index is $-\Gamma-1$; also,
see \citealt{Pozdnyakov_1983}).  Meanwhile, the shape of the red wing of
a relativistic line is given principally by the spin of the black hole,
but it is also affected\footnote{We note that
  increasing $q$ tends to decrease the line central peak, while
  increasing inclination has the opposite effect.} by the line emissivity $q$ and inclination $i$.

For reference, using canonical values $q=3$ and $i=60\degr$ 
  we find
the following correspondence between the shape of a red wing for a relativistic line (at a given $\spin$) and the downscattered wing of a narrow line due to Comptonization: $\spin = 1$
matches $\Gamma \approx 1.5$, $\spin = 0.7$ matches $\Gamma \approx
2.5$, $\spin = 0.4$ matches $\Gamma \approx 3.5$, and $\spin = 0$ matches
$\Gamma \gtrsim 4$.  It therefore follows that {\em if} 
Compton-scattering were being conflated with relativistic
distortion, this can be revealed though examination of data spanning a
range of $\Gamma$, or equivalently, a range of spectral
hardness. 

Ideally, one would examine {\em both} hard and soft (or
steep-power law; \citealt{RM06}) states to maximize this difference.
One expects that any bias introduced by this effect would lead a harder
spectrum to fit to a higher value of spin (or, equivalently, a
smaller inner radius).  We note that the most recent {\em NuSTAR} reflection studies of Cyg X--1 find $\spin = 0.93-0.96$ for the soft state and $\spin > 0.97$ in the hard state \citep{Walton_2016, Parker_2015}.  The direction of this mismatch in spin (corresponding a factor $\sim 25\%$ difference in $\rin$) is consistent with the type of bias one would expect from unmodeled coronal Comptonization.

We compare relativistic and Compton-scattering effects in
Figure~\ref{fig:mimic} in order to illustrate how the two may be
confused.  In the top panels, we present illustrative (simulated)
spectra in black comprised of both broad and narrow reflection
components (analyses using two reflection components being ubiquitous
for AGN, and commonplace for stellar BHs).  The broad component
originates in the regime of strong gravity while the narrow component is
produced from further away, where relativistic distortions are
negligible.  In red we illustrate that closely-matching spectra can be produced by
instead Compton-scattering a {\em single} narrow reflection component.  The left panels depict a representative
hard-state spectrum with spectral index $\Gamma=1.5$, and the right
panels a soft spectrum with $\Gamma=2.5$. (We note that many narrow-line Seyfert 1s show spectra with $\Gamma$ in this range, which may indicate that they have also have low $\fsc$.)
In the bottom panels, we show
as red dashed lines the shapes of the corresponding Compton-scattering profile for a narrow Fe line. For each, the resulting lineshape is comprised of sharp
transmitted and broad scattered components. 
As a reference, a relativistically
broadened Fe-line profile for a spin $\spin =0.9$ is overlaid as
a blue dotted line. Note the close correspondence between the
high-spin relativistic line and the Compton wing for the hard state in the
left-bottom panel; note also that the Compton-scattered lineshape for the soft state is not as broad as a high-spin (relativistic) line.   In both cases the upscattered power law from the Fe line contributes an excess blueward of the line-center, which is potentially detectable in a general case.  But since here the upscattered line lies  significantly below the Compton continuum in flux, the effect has minor impact in these examples.

In Figure~\ref{fig:rxscat} we show the net impact of coronal Comptonization on a full relativistic reflection spectrum (modeled via \relxill).  This  illustrates the generic case in which relativistic reflection features from an accreting black hole in an X-ray binary or AGN  undergo Compton-scattering through the corona.   The comingling of these effects applies the vast majority of X-ray reflection data of black holes, and motivates its presentation here. 
 In the figure, a green line shows the naked reflection emission from a nonrotating black hole ($\spin=0$), i.e., the reflection component  emitted by the disk prior to any coronal Compton scattering.  After passing through a corona with $\fsc=0.3$ and $\Gamma=2$, the scattered spectrum is both hardened and also smeared out in energy, with net result shown as the solid black curve.  Note for the black curve that the $\sim 30$~keV Compton hump and the red wing of the $\sim 6.7$~keV Fe line are broader, and that the reflection features appear muted relative to an enhanced (and harder) continuum.

\section{Towards a Composite Model}\label{section:fullmodel}  

\subsection{Model Components}
As the primary feature of our full spectral model, we ensure a
self-consistent linking of the fluxes between disk, Compton, and
reflection components.  Incorporating the effects of Comptonization by
the corona described above, we adopt two basic models which we will
employ in fitting \gx\ spectral data:

\noindent{\bf Model 1:} 

\noindent{\sc tbabs(simplcut$\otimes$(ezdiskbb+relxill)+xillver)},  

and

\noindent{\bf Model 2:}

\noindent{\sc tbabs(simplcut$\otimes$(ezdiskbb+relxill))}.   

Model~1 is a reformulation of the typical spectral model employed for
many supermassive and stellar black-hole systems (mentioned in
Section~\ref{subsec:downscatter}), which consists of both broad
(relativistic) and narrow reflection.  Here, the narrow reflection is
assumed to be produced far from the black hole, for instance in an outer
disk rim.  This distant emission would not undergo appreciable Compton
scattering by a central corona.  Because of this, we fix \xillver's setting for reflection fraction to -1 (the negative sign merely indicates that the continuum power-law is omitted), and fit for its normalization. Accordingly, we note that wherever a fit values of $\RF$ is shown, this refers specifically to the relativistic reflection component.
Model~2 is identical to Model~1 except that we
consider just a single relativistic reflection component.  Both include
a first-order treatment of reflection and Compton scattering (but not
higher order exchanges\footnote{For instance, we do not consider added
  contribution from reflection photons which Compton backscatter to
  re-illuminate the disk and produce further reflection, which again
  scatters, etc.  This contribution falls off at order $n$ as, roughly,
  $(\fsc\rf/(1+\rf))^n$.}).

Although omitted from the descriptions above, when we apply these models
to \gx\ we also include a Gaussian absorption line at $\sim 7.4$~keV
(this line may be an instrumental artifact rather than physical in origin, and is
discussed in great detail in G15\footnote{In G15, the line is positioned closer to 7.2~keV; there, 7.2~keV was determined in a fit to a larger data set which spanned a wide range of luminosities.  We are fitting just the highest luminosity subset -- Box A -- from that sample which reveals a modest increase in the line energy centroid when fitted independently (possibly related to the gas being more ionized).}.  {\sc tbabs} \citep{tbabs} describes
photoelectric absorption as X-rays traverse the line-of-sight
interstellar gas column with $N_{\rm H} = 5\times 10^{21} \cm^{-2}$
\citep{Felix_2016}, a quantity which we keep fixed throughout our
analysis.

{\sc ezdiskbb} is a spectral disk model\footnote{{\sc ezdiskbb} is a
  nonrelativistic disk model; nevertheless, it has one important
  advantage over its available relativistic counterparts {\sc bhspec}
  \citep{bhspec} or {\sc kerrbb} \citep{kerrbb}; namely, it can allow
  for disk truncation.  By contrast, the relativistic disk models make
  the assumption that $\rin=\risco$.} for a geometrically thin
multi-color accretion disk with zero torque at the inner-boundary
\citep{ezdiskbb}.  Importantly, {\sc ezdiskbb} has been variously shown
to recover essentially constant radii from soft black-hole spectra
(e.g., \citealt{Gou_2011, Chen_2016, Peris_2016}), demonstrating its utility here.

\relxill\ and \xillver\ are leading models of spectral reflection
\citep{xillver, relxill, relxill2}.  The essential difference is that
\relxill\ describes reflection from the inner-disk where gravitational
redshift and Doppler effects are important, and \xillver\ is used for
unblurred reflection occurring far from the relativistic domain.

In G15, we concluded that a composite model with a single reflection
component, akin to Model~2, did not adequately fit the bright hard
state of \gx.  Instead, a composite model akin to Model~1 (e.g., using
\relxill\ and \xillver\ together) was very successful.  The spectral
fits spanned a range of hard-state luminosities and together demanded
that the spin of \gx\ must be quite high ($\spin > 0.9$).  The same data
also exhibited a preference for quite modest disk truncation ($\rin \lesssim 5
R_g$) in the hard state, with the inner radius growing slightly larger at
lower luminosities.

\subsection{Parameter Constraints}

While the spectral fits in G15 were of high statistical quality, 
G15 identified three puzzling aspects of the best-fitting model: (1) The inner-disk reflection was associated
with a startlingly low reflection fraction ($\sim 0.2$), whereas values
closer to unity are expected\footnote{We note that the model definitions
  of normalization and reflection fraction in \relxill\ have been
  updated in more recent releases \citep{Dauser_2016}; $\rf$ using the
  updated definition (v0.4) is $\sim 0.3 \pm 0.03$.  While larger, this is still
  very far from unity.} (see, e.g., \citealt{Dauser_2014}).  (2) A very large Fe abundance, $>5$ times
solar, was demanded by the relativistic-reflection component.  At the
same time, the Fe abundance of the \xillver\ component was incompatible
with this large value and was instead consistent with a solar setting.
(3) Given the conclusion that the inner disk extends down to -- or very
near-to -- $\risco$, and given the high luminosity for these hard-state data, the lack of evidence for a thermal disk component in 
the PCA data is surprising.
	
Having incorporated the Compton scattering of inner-disk reflection into
our model, we now reexamine the most luminous data from G15.  We explore whether two distinct reflection components are now required (given the impact of including Comptonization), and then
go on to determine whether the fundamental conclusions of G15 hold up to this more holistic
approach. That is, we will test whether the lack of a detected disk
component can be reconciled with the high spin and modest-at-most
truncation of the disk in the bright hard state.  In addition, we test
whether \relxill\ alone provides a sufficient model for the reflection
signal, or whether the added \xillver\ component is still required.  As well, we consider whether the
three puzzles are in any way resolved using our
self-consistent model.

Parameters common to \relxill\ and \simplcut\ are tied to one another,
namely, inclination $i$, $\Gamma$, $\RF$, and $\ecut$.  For Model~1,
\xillver's ionization parameter is frozen to log~$\xi=0$ and the Fe abundance is
set to unity.  Only the normalization for \xillver\ is free (operationally, its setting for $\RF$ is fixed to -1).  As in G15, we keep the reflection
emissivity frozen to $q=3$ and fit for inner-disk radius while keeping
the spin at its maximum value $\spin=0.998$ (see Section~\ref{section:discussion}).  As a result of our
enforcing photon conservation, the normalization of \relxill\ is not
free.  
Instead, through comparison of the models, we fix the normalization to a fixed function of the disk and reflection parameters, changing in strength principally as $\RF$\ is adjusted.  Specifically, through exploration, we empirically derived a scaling relation that links the Compton power-law  emerging from disk (i.e., \simplcut$\otimes${\sc ezdiskbb}) to the illuminating power law (in \relxill\footnote{This is the power law obtained by setting $\RF = 0$ in the {\em unscattered} \relxill\ model.}) to within 5\% in flux. 
This serves as first-order approach to photon conservation, but again does not account for anisotropic redistribution in angle due to scattering or other geometric effects.

To anchor the {\sc ezdiskbb} temperature and normalization to values
corresponding to $\risco$, we turn to the abundant soft-state spectral
data of \gx\ whose Compton and reflection contributions are quite
minimal.  This is similar to the approach adopted by \citet{Kubota_2016}
in their study of a steep power-law state of \gx, which also employed a
thermal spectrum as a reference point.  We choose a $\sim 2\times10^6$
count spectrum from March 1, 1998 (ObsID 30168-01-01-00) which has an
X-ray flux comparable to that of the G15 Box~A data (see their Fig.~1), a factor $\sim 3$ below \gx's peak brightness.  The temperature
and normalization of the disk in question are, respectively $kT_{\rm
  soft} = 0.699~\keV$, and $N_{\rm soft} = 721$.  Owing to the abundant
evidence that soft-state disks reach $\risco$ (e.g.,
\citealt{Steiner_2010}), and that the stable disk radius can be
recovered in soft states to within several percent, we employ these
numbers as benchmark values for \gx's thermal disk at $\rin=\risco$.

We then allow for our hard-state spectrum to have a different $\rin$
which is free to take values $\rin \geq \risco$.  We link the thermal
disk properties to (i) the ratio $\rin/\risco$, which is a fit parameter
in \relxill, and also to (ii) the disk photon luminosity $\lph$ as
compared to our reference soft spectral state.

  In Appendix~\ref{append:disk}, we derive how the disk properties scale with changes from the photon luminosity and inner-radius.  The disk radiative efficiency $\eta \propto
\rin^{-1}$, and accordingly,  the seed thermal disk
in the hard state is described by

\begin{equation}\label{eq:tdisk}
kT_{\rm disk} = kT_{\rm soft} \left(\frac{\lph}{L_{\rm ph, soft}}\right)^{3/5} \left(\frac{\rin}{\risco}\right)^{-6/5}~\keV,
\end{equation}
and
\begin{equation}\label{eq:ndisk}
N_{\rm disk} = N_{\rm soft} \left(\frac{\lph}{L_{\rm ph, soft}}\right)^{-4/5} \left(\frac{\rin}{\risco}\right)^{18/5}.
\end{equation}

Equivalently, in terms of the mass accretion rate $\mdot$,
\begin{equation}\label{eq:tdisk2}
kT_{\rm disk} = kT_{\rm soft} \left(\frac{\mdot}{\mdot_{\rm soft}}\right)^{3/8} \left(\frac{\rin}{\risco}\right)^{-9/8}~\keV,
\end{equation}
and
\begin{equation}
\label{eq:ndisk2}
N_{\rm disk} = N_{\rm soft} \left(\frac{\mdot}{\mdot_{\rm soft}}\right)^{-1/2} \left(\frac{\rin}{\risco}\right)^{7/2}.
\end{equation}

\begin{figure*}[]
\begin{center}
  \includegraphics[width=2\columnwidth]{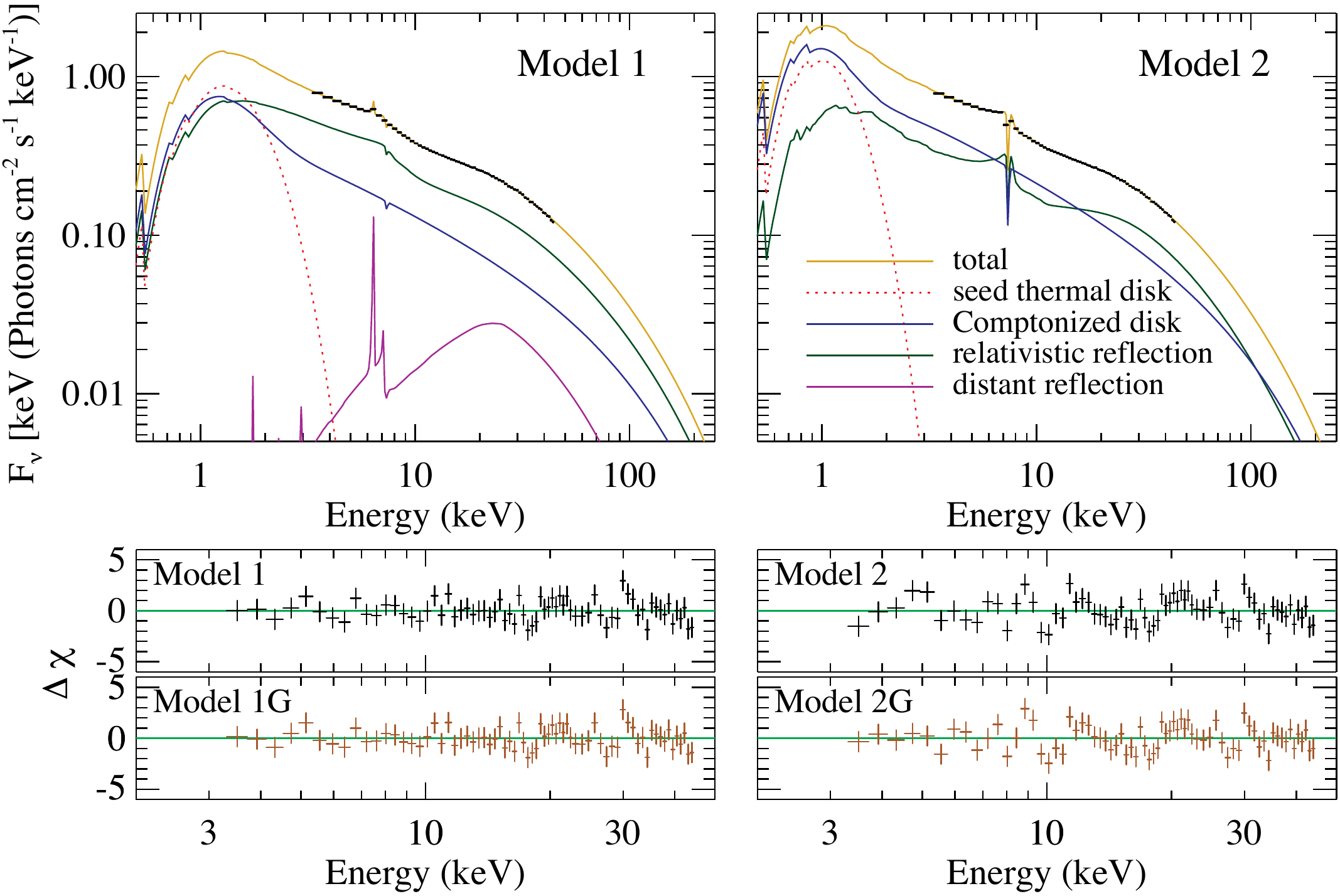}
  \caption{A fully self-consistent model comprised of a
    Compton-scattered thermal disk component and associated reflection emission, as applied to the bright hard-state data of GX 339--4.  The top panels
    show the composite fits and model components with the data in
    black.  Model 1 is shown in left panels and Model 2 in the right. The $\sim$7.4 keV absorption line is evident in each, but is much more pronounced in Model~2.
    Bottom panels show the fit residuals for both self-consistent and non-self-consistent variants of each model. Note the similarity in the residuals  among both variants for each of Models~1 and 2.  The associated fits are presented in
    Table~\ref{tab:results}.}\label{fig:spec}
\end{center}
\end{figure*}

\section{Results}\label{section:results}

\begin{deluxetable*}{lccccccl}
  \tabletypesize{\scriptsize}
  \tablecolumns{7}
  \tablewidth{0pc}
  \tablecaption{GX 339--4 spectral fit results}
 \tablehead{\colhead{Parameter} & \colhead{Prior\tablenotemark{a}} & \colhead{Model 1} & \colhead{Model 1G} & \colhead{Model 2} & \colhead{Model 2G} & }  
\startdata
$\Gamma$                        & F  & $         1.65^{+ 0.01}_{- 0.02} $   &   $              1.583 \pm 0.013 $    &     $   1.694 \pm 0.011        $  &  $       1.662^{+ 0.004}_{- 0.007} $  &   \\
$f_{\rm sc}$                     & F  & $                0.42 \pm 0.07 $    &                \nodata                &    $     0.32^{+ 0.09}_{- 0.18} $ &     \nodata          &    \\
$E_{\rm cut}$ (keV)             & LF  & $              93. \pm 4. $         &  $                    88. \pm 3. $    &      $     87. \pm 4.           $ &            $     90. \pm 3. $  &    \\
$i$ (deg.)                      & F  & $                 48.2 \pm 1.2 $     &     $               48.2 \pm 1.4 $    &   $    55.4^{+ 2.1}_{- 1.6}    $  &   $     58.6^{+ 0.8}_{- 1.1} $  &    \\
$R_{\rm in}/R_{\rm ISCO}(\spin=0.998)$   & F  & $ 1.52_{- 0.24}^{+ 0.18} $  &  $          1.6_{- 0.6}^{+ 0.2} $     &     $   5.3 \pm 0.8           $   &     $        1.4_{- 0.3}^{+ 0.1} $  &    \\
log $\xi$                       & F  & $         3.38^{+ 0.09}_{- 0.05} $   & $         3.37^{+ 0.08}_{- 0.05} $    &    $   2.783 \pm 0.017        $   &         $        2.803 \pm 0.021 $  &    \\
$A_{\rm Fe}$ (solar)             & LF & $            6.3^{+ 1.8}_{- 1.4} $  & $                 7.8 \pm 2.1 $       &     $   3.56 \pm 0.20          $  &           $       4.0 \pm 0.2 $ &              \\  
$\RF$                           & F  & $           0.8_{- 0.2}^{+ 0.4} $    & $       0.206^{+ 0.016}_{- 0.021} $   &    $  0.75_{- 0.08}^{+ 1.22}   $  &       $        0.492 \pm 0.018 $  &    \\
$N_{\rm relxill}$                 & d,LF &      $0.68 \pm 0.12  $           &   $                1.36 \pm 0.05 $    &      $0.88_{-0.20}^{+0.43}$       &            $     1.619 \pm 0.022 $  &    \\
$kT_{\rm disk}$ (keV)             &  d &               $0.31\pm0.03$        &                 \nodata               &        $0.15^{+ 0.05}_{- 0.02} $  &                  \nodata            &    \\  
$N_{\rm disk}$                    &  d &       $4300_{-1300}^{+2700} $      &                 \nodata               & $1.3^{+0.8}_{-0.6}\times10^{5} $  &                \nodata             &     \\
$N_{\rm xillver}$                 & LF & $                0.28 \pm 0.05 $   &  $                0.24 \pm 0.03 $     &            \nodata                &             \nodata           &     \\
$E_{\rm gabs}$ (keV)             & F  & $         7.38^{+ 0.16}_{- 0.12} $  &   $            7.39 \pm 0.11 $        &     $   7.37 \pm 0.04          $  &        $      7.45 \pm 0.03 $  &     \\ 
$\tau_{\rm gabs}$                & LF & $              0.018 \pm 0.005 $    &   $    0.019^{+ 0.006}_{- 0.004} $    &     $   0.154 \pm 0.019        $  &        $      0.143 \pm 0.015 $  &    \\
$\lph/L_{\rm ph, soft}$          & F  & $                0.55 \pm 0.10 $    &                 \nodata               &     $   2.1^{+ 1.5}_{- 0.5}    $  &  \nodata   \\                                              
$\mdot/\mdot_{\rm soft}$          & d & $            0.40 \pm 0.12   $      &                 \nodata               &     $   4.1^{+ 5.4}_{- 1.6}    $  &  \nodata   \\                                              
\hline
$\chi^2 / \nu$                    &   &                69.65 / 59           &                     68.21 / 59        &                      113.82 / 60  &      105.46 / 60  &
\enddata

\tablenotetext{a}{Priors are either flat (F) or flat on the log of
  the parameter (LF). Parameters marked ``d'' have values {\em derived} from the fit parameters.
For the self-consistent models, these
  values are not directly fitted for; instead, they are determined by
  fit parameters through the relationships outlined in
  Section~\ref{section:fullmodel}.}
\tablecomments{All fits were explored using MCMC; values and errors
  represent maximum posterior-probability density and minimum-width 90\%
  confidence intervals unless otherwise noted.}
\label{tab:results}
\end{deluxetable*}

For each of Models~1 and 2, we contrast our self-consistent analyses of \gx\ against the ``standard'' reflection formalism as adopted in G15, i.e., a model with no self-consistent linking of disk and reflection/Compton
components.  To highlight their tie to the G15 paper, these comparison ``standard'' models are termed Models~1G and 2G. 

In our analysis, we use XSPEC v12.9.0 \citep{XSPEC} to perform
a set of preliminary spectral fits. For consistency with G15, we employed
\relxill\ v0.2i, and we excluded the data in the first four 
channels while ignoring all data above 45~keV.  To ensure that the
redistribution of photons was accurately calculated, an extended
logarithmic energy grid was employed that sampled from 1 eV to 1 MeV at
1.4\% energy resolution via ``energies 0.001 1000. 1000 log''.
When a best fit was found, we estimate the errors through
a rigorous exploration of parameter space carried out using
Markov-Chain Monte Carlo (MCMC).  We employed the  {\sc python}
package {\sc emcee} \citep{emcee} as outlined in \citet{Steiner_lmcx1}, with further details provided in Appendix~\ref{append:mcmc}.

The corresponding results are summarized in Table~\ref{tab:results},
 We present our best fits for both self-consistent models in 
Figure~\ref{fig:spec}. The spectral components are shown as
solid colored lines and the data are shown in black. The seed disk
emission (i.e., the emission that would be observed from the bare disk
if the corona were transparent) is shown as a red dotted line. 
The sharp cosmetic absorption line near Fe is strong and pronounced in Model~2, but suppressed by the \xillver\ Fe-K$\beta$ edge in Model~1.  Still, its inclusion in Model~1 is significant at a $\sim 5~\sigma$ level.
 Each of Models~1 and 2 produce
{\em relatively} similar goodness-of-fits to their non-self-consistent counterparts and for both cases give nearly identical patterns
of residuals as shown in the bottom panels.  
 Thus, despite the constraints
which result from regulating the interplay of the spectral components, our self-consistent
framework very successfully models this extremely high-signal spectrum
of \gx\ in its peak bright hard state.

As a result of imposing self-consistency, the
low value of $\RF$ obtained in G15 (i.e., Models~1G and 2G) has increased to several times its
original value, now in a range close to unity which is aligned with expectation.
In part, this increase occurs because 
the appreciable Comptonization by the corona ($\fsc
\sim 0.4$) dilutes the equivalent width of the Fe line (and other
spectral features) since scattering a feature acts to blend it into the continuum (e.g., \citealt{Steiner_2016b}).  Accordingly, in order for the (scattered) model to match the Fe-line strength in the data, $\RF$ necessarily increases.  This same
effect was noted previously by \citet{Petrucci_2001}.  
For Model~1, surprisingly, the the best-fitting
$\rin$ {\em decreased} slightly relative to G15 (though within $\sim
1\sigma$ limits) by $\Delta\rin\approx0.1-0.2R_{\rm g}$.  However, for
Model~2 a new preferred solution emerges with an inner radius $\rin$ that is $\sim 3.8$ times
larger than for the non-self-consistent Model~2G.

  We find that 
self-consistency does not ameliorate the problem of high Fe abundance,
only slightly affecting its value.  And based on the significantly better quality-of-fit for the Model~1
compared to Model~2, we conclude that the new
self-consistent picture has {\em not} removed the need for two distinct
reflection components.  In fact, the $\chisq$ gap between the two best
fits is slightly larger for the self-consistent models. 

Apart from the question of which model flavor performs best, there are
several interesting differences between Models~1 and 2.  For Model~2,
because larger $\rin$ is preferred, the self-consistent model is matched with correspondingly higher $N_{\rm disk}$ and lower $kT_{\rm
  disk}$.  Model~2 also prefers lower values of the ionization parameter
(by a factor of $\sim$ 5) and of $\fsc$, but a higher inclination $i$.
Most striking, the $\mdot$ required for Model~1 is less than half that
of the reference, soft-state spectrum, whereas for Model~2, $\mdot$ is
instead {\em four}-times larger that for the corresponding soft state!  

Large $\rin$ resulting in large $\mdot$ is a
generic property of the self-consistent model because (from inspection of Eqns~\ref{eq:tdisk2} and
\ref{eq:ndisk2}) 
if $\rin$ is increased for a fixed $\mdot$, the photon flux diminishes and its peak moves to lower energy; both effects reduce the Compton power law's amplitude.  This is similar to the argument
presented by \citet{Dovciak_Done} who detailed how the luminosity of the
Compton component constrains the interplay between the seed photon flux
and the corona's covering factor. For a very luminous hard state (say, $\gtrsim 10\%$ of the Eddington limit),
large-scale truncation ($\rin >> \risco$) would imply highly super-Eddington values of
$\mdot$.

\begin{figure}[]
\begin{center}
  \includegraphics[width=1\columnwidth]{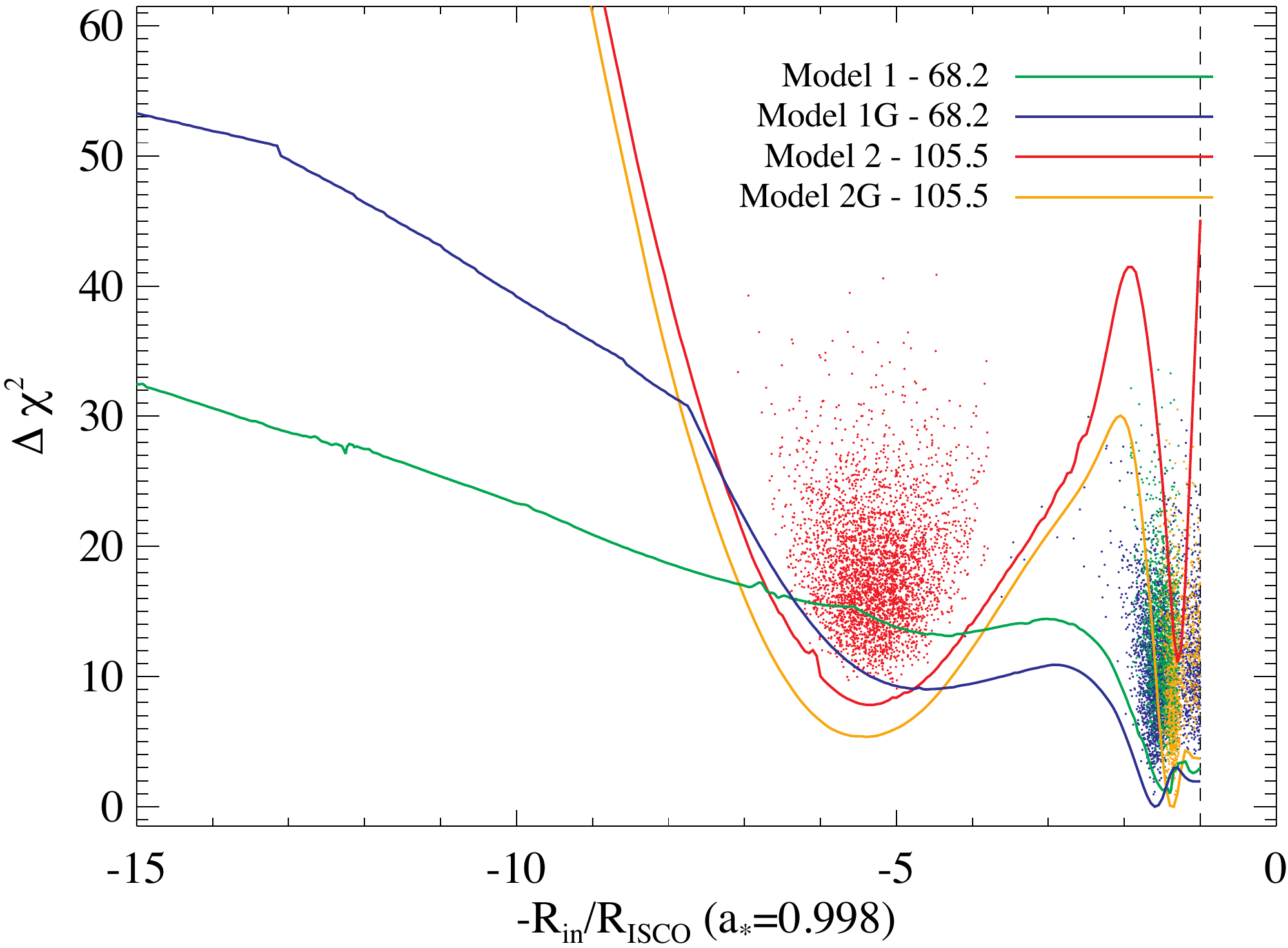}
  \caption{The impact of a self-consistent paradigm on the determination
    of the inner-disk radius.  We show the change in $\chi^2$ as
    the disk inner-radius is varied. The radius decreases to the right so
    that truncation is largest to the left. Model 1 is presented in
    green and Model~1G in blue; Model 2 is shown in
    red and Model~2G in orange.  The clouds of points show random draws from the MCMC chains.  Importantly, self-consistency penalizes small-radius
    solutions while reducing the penalty on truncated models. However,
    only for the disfavored Model 2 is the truncated disk preferred over
    the small-radius solution from G15 (i.e., Model~2G).  }\label{fig:steppar}
\end{center}
\end{figure}

To best examine the impact of the self-consistent model on the
inner-radius (or equivalently, on the resulting spin determination), we
present fit $\chisq$ versus $\rin$ in Figure~\ref{fig:steppar}.  
Here, solid lines show interpolations of ``steppar'' analyses
in \xspec\ (a routine which systematically optimizes the fit in sequential steps)
across a range of $\rin$.  Overlaid as point-clouds are random samplings
of the MCMC chains.  For Models 1 and 2, it is evident that imposing
self-consistency on the model has the effect of penalizing very low
values of $\rin$, and conversely making high $\rin$ solutions more
favorable.  This effect does not significantly alter the best-fitting
radius determined from Model~1, which significantly outperforms Model~2.
However, self-consistency does alter the fit landscape, which is a clear indication that the self-consistent constraints can impact results.

In Model~2, which is statistically disfavored, the effect of self-consistency is much more pronounced: The
$\chisq$ surface is double-troughed, with one minimum at $\rin \approx
1.4 \risco$ (=1.8~$R_g$) and another at a much larger radius of $\rin
\approx 5.3 \risco$ (=6.6~$R_g$). Thus, imposing self-consistency on
Model~2 drastically changes the solution, favoring substantial
truncation and penalizing mild truncation. The reason self-consistency
produces this result is two-fold: (1) The downscattered Compton line emission
contributes a flux excess in the same spectral region as the red wing of a 
relativistically-smeared line, and (2) the small-radius solution is
penalized for requiring a disk component that is not observed. As can be
gleaned from Figure~\ref{fig:spec}, Models~1 versus 2 could readily be
distinguished through direct detection of the thermal disk using a 
pile-up free detector with good low-energy sensitivity (e.g., {\it NICER}; \citealt{NICER}).

\section{Discussion}\label{section:discussion}

Our results differ with those of \citet{Wilkins_2015} concerning the
effects of Comptonization on reflection in the inner disk. These authors
find that strong Comptonization $\fsc \gtrsim 0.5$ has the effect of
{\em reducing} the breadth of line features emitted by the disk. The
difference arises from their view of the corona as being centrally
concentrated (though still porous) so that it very efficiently scatters
emission from the inner disk where the gravitationally-broadened red
wing is produced, while it is much less efficient at scattering emission
from further out in the disk. We note that there is mixed evidence for a
corona with this geometry; it is variously supported by microlensing
data of distant quasars (e.g., \citealt{Chartas_2009}) and reverberation
studies which require very compact corona \citep{Zoghbi_2010}, but it is
contested by \citet{Dovciak_Done} on the grounds that a corona so
compact would be starved of the seed photons required to produce the
observed X-ray luminosity.

The contrast between the results of \citet{Wilkins_2015} and ours
highlights the importance of understanding coronal geometry and the
necessity of investigating Compton-scattering effects for a range of
geometries in order to better understand systematics in employing
reflection spectroscopy to estimate disk radius and spin. At present,
these matters are all but unexamined.

\citet{Basak_Zdziarski} analyze \xmm\ spectra of \gx\ using a model
similar to ours, with the Compton, reflection and disk components
modeled using \nthcomp, \relxill\ and a standard multicolor disk.
While they cross-linked common parameters between the model components, a
significant shortcoming of their approach is not tying the disk seed
photons to the large number of photons scattered into the power law.
They also did not account for Compton scattering of reflection emission
in the corona. In contrast to our results, they found large-scale disk
truncation, with $\rin$ of tens of $R_g$, and they attributed a tentative detection of $\sim 0.2$~keV
thermal emission to the thermalization of power-law photons in the
disk's atmosphere.  Specifically, they claim $\rin$ to be an order of
magnitude larger than we find here.  We note that if such large values
of $\rin$ are correct, then to produce the observed $1-10~\keV$ bright
hard-state luminosity one would require the associated $\mdot$ to be extremely large, several times greater than the {\em peak} $\mdot$ of \gx's soft state\footnote{As determined by fixing $\rin$ to match their values when fitting the self-consistent implementations of Models~1 and 2}.  This is because for a given observed luminosity, when $\rin$ is increased the disk photon luminosity and peak temperature for a given $\mdot$ both diminish.  Both effects in turn cause the amplitude of the Compton power law to drop, and so match the data $\mdot$ must be large.  If one makes the reasonable assertion that the soft state peaks close to the Eddington limit, then large-scale truncation $R>>\Risco$ of a bright hard state would generally require a super-Eddington mass accretion rate.

\citet{Felix_2015} modeled low-luminosity hard-state \nustar\ spectra of
\gx. They found that the reflection component had a {\em harder}
power-law index $\Gamma$ than the direct continuum component.  The modest
difference in the spectral indexes $\Delta\Gamma\approx0.2-0.3$ is
plausibly explained in our model by the hardening of the reflection
spectrum due to Compton scattering in the corona. This can be readily seen in the dashed line of Fig.~\ref{fig:rxscat} which is harder than the input $\Gamma=2$.  Although difficult to
discern by eye in Figure~\ref{fig:spec}, the Compton-scattered
portion of the reflection component emerges from the corona with a harder spectral index, $\Delta\Gamma\approx0.15$. This is because, like the Comptonized thermal photons, the Comptonized reflection spectrum is boosted in energy when scattering in the corona.  As
one would expect, the magnitude of $\Delta\Gamma$ can be shown to grow with
$\fsc$. This hardening explains why the
reflection component can be brighter than the Compton continuum even if
$\RF<1$, as it quite apparently is for Model~1 in
Figure~\ref{fig:spec}. This occurs because the Compton-scattered
reflection is effectively boosted by an additional factor of the
``Compton y'' parameter.

In applying our reflection model, we have proceeded under the assumption
of maximal spin ($\spin = 0.998$), which effectively sets $\risco$ for
the thermal state data.  If we had instead assumed any lower value for
spin, the tension between the disk component (given that the data rule out a bright, hot disk) and reflection component (which prefers a disk proximate to  the horizon) would have
strictly increased, and the fits would have worsened.  In this sense, our
approach of adopting maximal spin provides a conservative estimate of
the importance of self-consistency on the modeling.


\section{Conclusions}\label{section:concs}

In summary, we have demonstrated that the Comptonization of the Fe 
line and other reflection features in a hot corona can mimic the effects
of relativistic distortion -- and potentially affect estimates of black
hole spin -- by producing a downscattered red wing.  This notionally
calls into question the prevalence of dual-reflection component spectral
models in which one component is in the relativistic domain, and hence
strongly blurred, while the other one is assumed to occur far from the
black hole and to be correspondingly narrow. The precise shape of a
relativistically-distorted line is strongly affected by the spectral
index, and hence it changes as a source transitions between hard and
soft states. The Comptonization of reflection features is particularly
important for hard spectral states because they are dominated by
Compton-scattered photons from the thermal disk and their byproduct --
 reflection. Given that thermal disk photons are strongly
Comptonized in hard states, the
associated reflection emission is inevitably strongly Comptonized as
well.

We have incorporated reflection and Comptonization into a
self-consistent disk-coronal spectral model that properly conserves
photons between the thermal, reflection, and Compton components. Within
this framework, and using two specific models (Models 1 and 2), we
analyzed a bright hard-state \rxte\ spectrum of \gx\ containing
44-million counts, which was analyzed previously by G15.  An important
constraint was obtained using a soft (thermal dominant) state
spectrum of \gx\, at approximately the same luminosity as the G15 data to anchor $\risco$ and the hard-state disk scaling. From our analysis using the self-consistent
models, we find that a single Compton-scattered reflection component is not preferred for \gx; as in G15, the dual-reflection model is strongly 
favored.  In fact, imposing self-consistency on the dual-reflection model and including the
effects of Compton scattering on the broad reflection component has minor effect on the reflection best-fitting parameters aside from the reflection fraction
$\RF$ which is larger by a factor $\sim 2-4$. At the same time and
importantly, the large change in the inner-disk radius for the
disfavored Model~2 illustrates that self-consistency has the potential
to significantly affect reflection estimates of black hole spin.


Additionally, we find that {\em bright hard states} cannot be reconciled with
{\em large-scale} disk truncation unless either (1) the mass accretion rate
$\mdot$ is super-Eddington (well in excess of the peak soft-state
$\mdot$) or else (2) the hard state power-law component is attributed to
another radiation mechanism. This conclusion is a consequence of the
photon conservation that is a bedrock of our model. For a
fixed $\mdot$ one can intuitively understand this result: a disk with
large $\rin$ is cool, produces few photons, and has a Compton component
that peaks at lower energy. At the same time, the corona cannot be very
small, or else it would require an enormously luminous disk to scatter
sufficient photons into the observed power-law component.

For a multi-epoch study of a
source, the effects of Compton scattering on reflection features can be
assessed as the source ranges over soft and hard states; i.e., as the
photon index $\Gamma$ varies.  Our model predicts that any bias related to Compton scattering of the reflection emission having been omitted from modeling efforts should cause soft-state fits to measure lower spins than found in hard states for the same source (e.g., consistent with the findings of the most recent {\em NuSTAR} reflection studies of Cyg X--1).

As illustrated in Figure~\ref{fig:steppar}, enforcing self consistency
has the potential to very significantly affect estimates of the inner
disk radius and spin when modeling a reflection spectrum in the presence
of a thermal component. Furthermore, this approach provides new, mutual
constraints, e.g., on $\rin$ and $\mdot$. Self-consistent reflection
models therefore deserve exploration and further development.

Going forward, given that the power-law component in hard states is
strongly Comptonized, standard practice should include 
Comptonization of the reflection component in the corona, particularly for studies
aimed at constraining $\rin$ or $\spin$. 

\acknowledgments

JFS has been supported by NASA Einstein Fellowship grant PF5-160144.  We thank Ramesh Narayan and Charith Peris for helpful discussions, and the anonymous referee.

\begin{appendix}
\section{simplcomp}\label{append:simplcomp}

In addition to our adopted scattering kernel, which computes the photon redistribution following an exponentially cutoff power law, we have also implemented a kernel ``{\em comp}'' using the Comptonization model \nthcomp\ to
numerically compute the scattered distribution at each $E_0$.  In \xspec, toggling the energy turnover parameter positive or negative
switches between kernels. A positive value calls {\em comp}
  \citep{nthcomp2, nthcomp} and the value sets the electron temperature
  $kT_e$, whereas a negative value invokes Eqn.~\ref{eqn:simplcut} and
  the absolute value of its setting gives $\ecut$.

As evident in Figure~\ref{fig:comptonization}, the kernels produce nearly identical profiles for the downscattered
component, whereas the upscattered spectral shape is more sharply curved at its turnover
using the physically rigorous {\em comp} kernel.  We compare the fitting results obtained with both kernels in Table~\ref{tab:append}.  Aside from the large value of $\ecut/kT_e \sim 4-6$, the kernels produce similar results.
The spectral index is slightly larger for the {\em comp} versus $\Ecut$ kernels for both Models~1 and 2.  This is because the $\Ecut$ kernel {\em overestimates} the power-law curvature at energies well below the cutoff (as can be gleaned through close comparison between the kernel shapes in Fig.~\ref{fig:comptonization} at energies $\sim 2-50$~keV).  This means that for the $\ecut$ and {\em comp} kernels to match over the intermediate energy range below the turnover, an inherently steeper index is needed with {\em comp}.

Beyond the fits in Section~\ref{section:results}, which aims to provide 
ready comparison with the results of G15, we have also
explored the use of a new beta-version of \relxill\ code which employs 
a \nthcomp\ input continuum for reflection computations, as opposed to
the cutoff power law. When applied in conjunction with the {\em comp}
scattering kernel, this more physical continuum yields fits for each of
Models~1 and 2 that are better than their counterparts in
Tables~\ref{tab:results} and \ref{tab:append}.  This bolsters the expectation that a physically accurate
Comptonization model is preferred.  A description of the \nthcomp\
continuum implementation in \relxill\ will be presented in future work.

\begin{deluxetable*}{lcccccccl}
  \tabletypesize{\scriptsize}
  \tablecolumns{8}
  \tablewidth{0pc}
  \tablecaption{GX 339--4 spectral fit comparison between $\ecut$ and {\em comp} kernels}
 \tablehead{\colhead{Parameter} & \colhead{Model 1 ($E_{\rm cut}$)} &  \colhead{Model 1 (comp)}  & \hspace{1cm}  & \colhead{Model 2 ($E_{\rm cut}$)} & \colhead{Model 2 (comp)} & }  
\startdata
$\Gamma$                            &       $         1.65^{+ 0.01}_{- 0.02} $  &     $             1.708 \pm 0.015 $   &  &   $   1.694 \pm 0.011        $      & $  1.767^{+ 0.006}_{- 0.013} $  & \\
$f_{\rm sc}$                        &       $         0.42 \pm 0.07 $           &     $               0.44 \pm 0.06 $   &  &  $     0.32^{+ 0.09}_{- 0.18} $     & $  0.35^{+ 0.12}_{- 0.18}    $  & \\
$E_{\rm cut} {\rm ~or~} kT_e$\tablenotemark{a} (keV)  & $     93. \pm 4. $      &     $          18.7^{+ 1.4}_{- 1.0} $ &  &    $     87. \pm 4.           $     & $  19.9 \pm 0.8            $  &   \\
$i$ (deg.)                          &  $                 48.2 \pm 1.2 $         &     $                48.6 \pm 1.2 $   &  & $    55.4^{+ 2.1}_{- 1.6}    $      & $  54.7 \pm 1.4            $  &   \\
$R_{\rm in}/R_{\rm ISCO}(\spin=0.998)$  &       $   1.52_{- 0.24}^{+ 0.18} $    &     $       1.40_{- 0.13}^{+ 0.17} $  &  &   $   5.3 \pm 0.8           $       & $  5.0 \pm 0.7            $  &    \\
log $\xi$                            & $         3.38^{+ 0.09}_{- 0.05} $       &     $        3.51^{+ 0.14}_{- 0.09} $ &  &  $   2.783 \pm 0.017        $       & $  2.776^{+ 0.019}_{- 0.014} $  & \\
$A_{\rm Fe}$ (solar)                 &  $            6.3^{+ 1.8}_{- 1.4} $      &     $           9.7^{+ 0.2}_{- 1.5} $ &  &   $   3.56 \pm 0.20          $      & $  4.6 \pm 0.3             $  &   \\
$\RF$                                & $           0.8_{- 0.2}^{+ 0.4} $        &     $          0.8_{- 0.1}^{+ 0.5} $  &  &  $  0.75_{- 0.08}^{+ 1.22}   $      & $  1.1_{- 0.4}^{+ 1.3}      $  &  \\
$N_{\rm relxill}$\tablenotemark{b}   &      $0.68 \pm 0.12  $                   &       $0.55^{+ 0.17}_{- 0.04} $       &  &    $0.88_{-0.20}^{+0.43}$           &   $0.78_{-0.21}^{+0.45}$        & \\
$kT_{\rm disk}$\tablenotemark{b} (keV)  &                $0.31\pm0.03$          &      $0.31\pm0.03$                    &  &      $0.15^{+ 0.05}_{- 0.02} $      & $0.16^{+ 0.04}_{- 0.02} $     &   \\
$N_{\rm disk}$\tablenotemark{b}      &       $4300_{-1300}^{+2700} $            &    $ 4200_{-1100}^{+1400}$            &  & $1.3^{+ 0.8}_{- 0.6} \times 10^{5} $ & $1.1^{+ 0.7}_{- 0.4} \times 10^{5} $ & \\
$N_{\rm xillver}$                    & $                0.28 \pm 0.05 $         &     $                0.26 \pm 0.07 $  &  &          \nodata                    &       \nodata                  &  \\
$E_{\rm gabs}$ (keV)                 & $         7.38^{+ 0.16}_{- 0.12} $       &     $   7.35^{+ 0.13}_{- 0.06} $      &  &   $   7.37 \pm 0.04          $      & $  7.34 \pm 0.03            $  &  \\
$\tau_{\rm gabs}$                    & $              0.018 \pm 0.005 $         &     $              0.030 \pm 0.006 $  &  &   $   0.154 \pm 0.019        $      & $  0.174 \pm 0.018          $  &  \\
$\lph/L_{\rm ph, soft}$         & $                0.55 \pm 0.10 $              &     $     0.48^{+ 0.12}_{- 0.04} $    &  &   $   2.1^{+ 1.5}_{- 0.5}      $    & $  1.9^{+ 1.3}_{- 0.4}        $  & \\
$\mdot/\mdot_{\rm soft}$\tablenotemark{b}        & $                0.40 \pm 0.12   $            &     $     0.34^{+ 0.12}_{- 0.06} $    &  &   $   4.1^{+ 5.4}_{- 1.6}      $    & $  3.3^{+ 4.3}_{- 1.1}        $  & \\
\hline
$\chi^2 / \nu$                  &                69.65 / 59                     &                      73.16 / 59       &  &                    113.82 / 60      &            121.39/ 60          &  
\enddata

\tablenotetext{a}{A comparison between $\kte$ and $\Gamma$ for a
  particular geometry can be used to infer the equivalent corona optical
  depth $\tau$.  Using {\sc compTT} \citep{comptt}, for slab and
  spherical coronal geometries, the equivalent $\tau$ ranges from $\sim
  2-5$, respectively.  By comparison, $\fsc$ is much lower, which might
  indicate either a low covering factor or a high porosity due to a clumpy
  corona.}
\tablenotetext{b}{These
  values are not directly fitted for; instead, they are determined by
  fit parameters through the relationships outlined in
  Section~\ref{section:fullmodel}.}
\tablecomments{All fits were explored using MCMC; values and errors
  represent maximum posterior-probability density and minimum-width 90\%
  confidence intervals unless otherwise noted.}
\label{tab:append}
\end{deluxetable*}

\section{Deriving the disk's scaling laws}\label{append:disk}

Here, we present a concise derivation of the scaling laws employed for our self-consistent models, which relate the disk's color temperature $T_{\rm col}$ and normalization $N$ to the disk inner radius $R_{\rm in}$ and photon luminosity $L_{\rm ph}$.  Our prescription assumes that the hard state accretion disk is described by a standard thin-disk model (e.g., \citealt{SS73}) down to $\rin$.  $\rin$ can be as small as $\risco$, but if it is large then inward of $\rin$ the disk is assumed to transition into a radiatively inefficient, geometrically thick and optically thin flow (e.g., as in the sequence described by \citealt{Esin_1997}).  

Allowing that the soft-state disk extends down to $R_{\rm in, soft} = \risco$, the ratio of hard-state and soft-state luminosities depends purely on the luminosity of each, and their respective efficiencies, $\eta \propto R^{-1}$.  Recall that color temperature $T_{\rm col} = f T_{\rm eff}$ where $f \propto T_{\rm col}^{1/3}$, so $T_{\rm eff} \propto T_{\rm col}^{2/3}$.
We adopt a simplifying notation where the lowercase letter indicates a dimensionless scaling for the hard-to-soft ratio.  Specifically, $l \equiv \frac{L_{\rm hard}}{L_{\rm soft}}$, $r_{\rm in} \equiv \frac{R_{\rm in, hard}}{\risco}$, $\dot{m} \equiv \frac{\mdot_{\rm hard}}{\mdot_{\rm soft}}$, and $t \equiv \frac{T_{\rm hard}}{T_{\rm soft}}$.   The luminosity,

\begin{equation}
l = \dot{m} r_{\rm in}^{-1} = r_{\rm in}^2 t_{\rm eff}^4.
\label{eq:lum}
\end{equation}

The photon luminosity scales simply as the energy luminosity divided by the color temperature of the disk, 

\begin{equation}
l_{\rm ph} = r_{\rm in}^2 t_{\rm eff}^4 t_{\rm col}^{-1} =  r_{\rm in}^2 t_{\rm col}^{5/3}. \\
\label{eq:lph}
\end{equation}

Equivalently, this yields a temperature scaling of 
\begin{equation}
t_{\rm col} = l_{\rm ph}^{3/5} r_{\rm in}^{-6/5}.
\label{eq:tscale}
\end{equation}

From \citet{ezdiskbb}, the disk normalization $N$ scales as $R_{\rm in}^2/f^4$.  Here, we substitute the color temperature dependence of $f$, and define $n\equiv N_{\rm hard}/N_{\rm soft}$.  Then,
\begin{equation}
n = r_{\rm in}^2 t_{\rm col}^{-4/3}.
\end{equation}

Substituting the temperature scaling of Eqn~\ref{eq:tscale}, we obtain the scaling relation for normalization,
\begin{equation}
n = l_{\rm ph}^{-4/5} r_{\rm in}^{18/5}.
\label{eq:nscale}
\end{equation}

Eqns~\ref{eq:tscale} and \ref{eq:nscale} correspond to Eqns~\ref{eq:tdisk} and \ref{eq:ndisk} in Section~\ref{section:results}.

While the photon luminosity is a natural quantity for the examination of Comptonization, the more natural description of accretion employs $\mdot$.  Combining Eqns~\ref{eq:lum} and \ref{eq:lph} one finds,
\begin{equation}
l_{\rm ph} = \dot{m} t_{\rm col}^{-1} r_{\rm in}^{-1} = \dot{m}^{5/8} r_{\rm in}^{1/8}.
\label{eq:lph2}
\end{equation}

Substituting Eqn~\ref{eq:lph2} into Eqns~\ref{eq:tscale} and ~\ref{eq:nscale}, one obtains
\begin{eqnarray}
t_{\rm col} = \dot{m}^{3/8} r_{\rm in}^{-9/8}, \\
n           = \dot{m}^{-1/2} r_{\rm in}^{7/2}.
\label{eqns:mdotscale}
\end{eqnarray}

\section{MCMC run setup}\label{append:mcmc}
The python {\sc  emcee} algorithm distributes a set of ``walkers'' that together explore 
the parameter landscape in a sequence of affine-invariant
``stretch-move'' steps.  For our models, we used 50 walkers which were
initially scattered about the point of the best fit and run for between
500,000 and 1,000,000 steps apiece.  Our threshold for convergence was
twenty auto-correlation lengths per walker in {\em each} parameter,
corresponding to a minimum of 1,000 effective samplings.  The associated
computations were parallelized and the runs required approximately 2
CPU-years in aggregate.  In the fits, as described in
\citet{Steiner_lmcx1}, each parameter was remapped using a logit
transformation to regularize the space over which it was sampled from a
finite interval to the real line, and each parameter was assigned a
noninformative prior which was flat on either the parameter value, or on
the logarithm of its value (for scale-independence).

\end{appendix}

\end{document}